# Analog Noise Reduction in Enzymatic Logic Gates

**Dmitriy Melnikov**, **Guinevere Strack**, **Marcos Pita**, **Vladimir Privman\*** and **Evgeny Katz\***

*Department of Chemistry and Biomolecular Science, and Department of Physics,*
*Clarkson University, Potsdam, NY 13676*

**\***Corresponding authors:

E-mail: ekatz@clarkson.edu, phone: +1-315-268-4421, fax: +1-315-268-6610 (E. Katz)

E-mail: privman@clarkson.edu, phone: +1-315-268-3891, fax: +1-315-268-6610 (V. Privman)

## Abstract

In this work we demonstrate both experimentally and theoretically that the analog noise generation by a single enzymatic logic gate can be dramatically reduced to yield gate operation with virtually no input noise amplification. This is achieved by exploiting the enzyme's specificity when using a co-substrate that has a much lower affinity than the primary substrate. Under these conditions, we obtain a negligible increase in the noise output from the logic gate as compared to the input noise level. Experimental realizations of the **AND** logic gate with the enzyme horseradish peroxidase using hydrogen peroxide and two different co-substrates, 2,2'-azino-*bis*(3-ethylbenzthiazoline-6-sulphonic acid) (ABTS) and ferrocyanide, with vastly different rate constants confirmed our general theoretical conclusions.

Link to future updates of this article and to manuscript with higher-resolution images

## 1. Introduction

Recently there has been a significant interest in chemical[1] and biochemical[2] logic and computing. At present, there are many chemical-computing studies that have relied on molecular systems to mimic functions typical for electronic computing devices such as simple Boolean logic gates, **AND**,[3] **OR**,[4] **XOR**,[5] **NOR**,[6] **NAND**,[7] **INHIB**,[8] **XNOR**,[9] as well as more complex operations.[10] Chemical systems are in principle capable of performing computations at the level of a single molecule[11] resulting in nanoscaling of the computing units.[12] One of the most important challenges in chemical and biochemical computing is networking and its scaling up which requires exploration of fault tolerant design of complex multi-reactant systems.[1,10,13–17]

Biochemical computing and logic systems utilize biomolecules, such as proteins/enzymes,[18] DNA,[19] RNA,[20] and whole biological cells[21] to process information. Enzyme-based biomolecular systems have the advantage of specificity (enzymes are naturally very selective in their chemical functions) and usability in complex "chemical soup" environments. Most reactions in a living cell are controlled (catalyzed) by specific enzymes, creating a complex network of interconnected biochemical processes. This means that in principle, enzyme-based information processing units can be made highly scalable giving rise to artificial biocomputing networks performing various logic functions and mimicking natural biochemical pathways.[22] Compared with electronic counterparts, biomolecular computing systems also have the advantage of being able to process biochemical information in the form of chemical signals directly from biological systems. This is important for interfacing of the resulting bioelectronic "devices" with living organisms, for potential biomedical applications.[17,23]

Recent experiments in enzymatic biocomputing demonstrated realization of single Boolean gates[14,15,24] as well as networking of several (up to 3-4 presently) gates.[16,17,25] Similar logic operations were also realized using non-biochemical systems,[1] but biochemical variants offer greater simplicity of the assembled logic schemes. With increasing number of elements in enzyme-based networks, exploration of noise suppression approaches becomes necessary.[14–16] Fault tolerant (error free) information processing in small enzymatic networks can be accomplished by individual logic gate optimization for the reduction of the "analog" noise build-up[15] and/or by network topology.[16] For even larger networks, another, "digital" mechanism[14–16] of noise amplification emerges. It is combated by redundancy in network design and requires appropriate new network elements for filtering, rectification, etc.

In this work we address aspects of optimizing a single enzymatic logic gate with the purpose of the analog noise reduction. Our experimental approach is summarized in Section 2. Section 3 presents the theoretical approach for describing the gate function in terms of the "Boolean logic" input values (**0,0**; **0,1**; **1,0**; **1,1**), yielding the Boolean **AND** gate output values **1** (for input **1,1**) or **0** (for inputs **0,0**; **0,1** or **1,0**). Achieving low noise amplification (and ultimately, noise reduction i.e., filter-like) functioning near all the "logic" input values (**0,0**; **0,1**; **1,0**; **1,1**) for a single gate with enzyme(s) as the gate "machinery" has been an elusive goal. Earlier successes have been for a gate with enzymes used instead as the inputs,[15] as well as by optimizing the relative activity of enzyme-based gates within a three-gate network.[16]

The challenge has been several-fold, as alluded to in our previous study.[15] Since the reactant concentrations are normalized by the reference values to define the "logic" variables (see Section 3), then in terms of these variables the dependence of the kinetics on the catalyst (enzyme) activity turns out rather weak. Thus, experimentally, in order to adjust the gate function, large variations in the enzyme activity are required which might not be practical to achieve by changing only the enzyme concentration. In addition, the theoretical analysis of the logic-gate function has to be limited to a few-parameter shape fitting of the output vs. input response function in terms of simple, irreversible substrate-reaction rate equations, as illustrated shortly. Indeed, the data obtained by experimentally mapping out the gate response, are not sufficiently sensitive for a multi-parameter fit to describe the detailed enzymatic kinetics which is usually multi-pathway type, with most of the elementary reactions reversible.

Specifically, here we consider a gate with two inputs, one of which is the co-substrate to be specified later (since we consider two different choices), the concentration of which will be denoted $I_1$. It is oxidized in the enzymatic process which also takes in the substrate, hydrogen peroxide ($H_2O_2$) which is the second logical input, $I_2 = [H_2O_2]$. This reaction is catalyzed by the enzyme horseradish peroxidase (HRP), at concentration $E$. As described above and in our previous study,[15] for logic-gate function parameterization we have to use a simplified (approximate) kinetic description, here with two irreversible steps, assuming a single intermediate complex of concentration $C$,

$$E + I_2 \xrightarrow{r} C, \qquad (1)$$

$$C + I_1 \xrightarrow{R} P + \cdots, \qquad (2)$$

where $P$ is the concentration of one of the two products (the oxidized co-substrate) which is detected as the logic-gate outputs, while $r$ and $R$ are the effective rate constants for the reactions. We note that for the regime of relevance to our experiments, the reduction of $H_2O_2$ catalyzed by HRP is typically in the diffusion-limited regime and can be here approximated as irreversible,[26] with a single, large rate constant, $r$. The main approximation, made to reduce the number of adjustable parameters, in our kinetic scheme is in fact that of lumping several possible pathways involving variants of the complexes formed, with at least two complex-formation steps,[27] into the simple "complex" entity of concentration $C$. The corresponding rate equations are

$$\frac{dI_2(t)}{dt} = -r[E(0) - C(t)]I_2(t), \qquad (3)$$

$$\frac{dC(t)}{dt} = r[E(0) - C(t)]I_2(t) - RC(t)I_1(t), \qquad (4)$$

$$\frac{dP(t)}{dt} = RC(t)I_1(t), \qquad (5)$$

where for clarity we explicitly showed all the time dependences and we used the relation $E(t) = E(0) - C(t)$ to express all the time $t > 0$ quantities in terms of the three input-output chemicals.

Note that initially $P(0), C(0) = 0$, but the concentrations $E(0), I_{1,2}(0)$ can vary depending on the environment in which the enzymatic gate is used, for the inputs, $I_{1,2}(0)$, and on our selection of the gate "activity" via $E(0)$. Thus, even with the present extremely oversimplified modeling of the enzymatic kinetics of HRP, we have several parameters to deal with, in connection with fitting the gate function and attempting to optimize it, as further addressed in Section 3. Direct optimization has proven elusive, primarily due to limited sensitivity of the shape of the response function in terms of the logic variables to the value of $E(0)$.[15,16] Instead, ideas such as using enzymes with self-promoter substrates, typical of many allosteric enzymes,[22] as well as other systems involving additional kinetic processes,[15,28] all yielding "sigmoid" response functions, were proposed, but thus far were not experimentally realized, though sigmoid response is common in natural cell processes.[28,29]

The main finding of the present work is a new mode of single gate optimization which does not attempt to use more complicated chemical reaction kinetics. We have identified a regime of functioning of a single enzyme with an unusual shape of the logic-variable response surface (see Sections 3-4 for definitions and details), which can yield logic gates with practically no analog noise amplification. As mentioned in connection with equation (1), the rate constant $r$ is large.[26] On the other hand, the rate constant $R$ describing the oxidation of the co-substrate, $I_1$, can be varied over a broad range of values as it is known to depend on the choice of the chemical co-substrate.[26] It turns out that the ratio between $r$ and $R$ dramatically affects the degree of amplification of the analog noise generated by this gate, and, for sufficiently large $r/R$, yields a new response-surface shape with desirable low noise-amplification properties. We used 2,2'-azino-*bis*(3-ethylbenzothiazoline-6-sulphonic acid) (ABTS) as the "fast" co-substrate, and $K_4Fe(CN)_6$ (ferrocyanide) as the "slow" co-substrate to experimentally confirm these observations. Our results are presented in Section 4, followed by a concluding discussion in Section 5.

## 2. Experimental

### 2.1. Chemicals and reagents

The biocatalyst and some other chemicals were purchased from Sigma-Aldrich and used as supplied: horseradish peroxidase type VI (E.C. 1.1.1.7) (HRP), 30% hydrogen peroxide, 2,2'-azino-*bis*(3-ethylbenzothiazoline-6-sulphonic acid) (ABTS). From T. J. Baker, we used sodium sulfate, anhydrous powder, 99%. From Fisher Scientific, potassium ferrocyanide, 98.5%, potassium nitrate, 99.5% ACS certified ($KNO_3$), and citric acid anhydrous, certified (ACS).

Ultrapure water (18.2 MΩ/cm) from NANOpure Diamond (Barnstead) source was used in all of the experiments.

## 2.2. The chemical logic gate and input signals

The ABTS-based **AND** gate was carried out in an aqueous solution consisting of $H_2O_2$ ranging from 0.05 μM to 50 μM, ABTS ranging from 0.01 mM to 0.5 mM, and 0.5 nM HRP in 25 mM McIlvaine citrate-phosphate buffer, pH = 5.0. The HRP concentration was determined spectrophotometrically[30] using $\varepsilon_{403} = 102$ mM$^{-1}$cm$^{-1}$. The ferrocyanide-based **AND** gate was composed of $H_2O_2$ ranging from 0.05 mM to 0.25 mM, $K_4Fe(CN)_6$ ranging from 0.01 mM to 1.5 mM, and 0.05 μM HRP in 10 mM phosphate buffer, pH = 6.0. The ionic strength of the reaction mixture was maintained at 0.11 M. The phosphate buffer contributed to 0.01 M of the ionic strength, while ferrocyanide and $KNO_3$ contributed to the remainder. Ferrocyanide stock solutions were made within 1 hour of usage and protected from light to prevent photodecomposition.

Hydrogen peroxide and ABTS were initially used as the chemical input signals to activate the **AND** gate. Based on the theoretical analysis, the **AND** gate was subsequently optimized by replacing ABTS with ferrocyanide. The absence or presence at a particular reference concentration, of each chemical input was considered as input signal values **0** or **1**, respectively. However, in order to map out the "response surface," the concentrations of the chemicals operating as input signals were varied for values between **0** and the reference Boolean logic-**1** concentrations, as detailed in Sections 3 and 4.

## 2.3. Measurements

The measurements were performed using a UV-2401PC/2501PC UV-visible spectrophotometer (Shimadzu, Tokyo, Japan) at 25 ± 2 °C. The reaction took place in a 1 mL polyacrylamide cuvette. Upon addition of chemical inputs in different combinations, the change in absorbance, $\Delta A$, was measured at $\lambda = 414$ nm in the case of ABTS$_{ox}$ (where the subscript denotes "oxidized") and $\lambda = 425$ nm for ferricyanide. The absorbance values were converted to molar concentrations using $\varepsilon_{414} = 36 \text{ mM}^{-1}\text{cm}^{-1}$ for ABTS[31] and $\varepsilon_{425} = 1.06 \text{ mM}^{-1}\text{cm}^{-1}$ for ferricyanide.[26] Here we consider results for the reaction time ("gate time;" see Section 3) of 60 sec, but results for another time (120 sec) are commented on in Section 5.

## 3. Theoretical Approach

For the **AND** gate function, we set Boolean **0** (logic-**0**) as zero initial concentration of the input chemicals (one or both), $I_{1,2}(0)$, and the output, $P(t)$, at the "gate time," $t = t^{\text{gate}}$. The Boolean **1** (logic-**1**) inputs correspond to reference concentrations of $I_1^{\text{gate}}$ and $I_2^{\text{gate}}$ at time 0, and together with the gate time, $t^{\text{gate}}$, in our case were selected as experimentally convenient values, but in applications will be set by the gate environment or by the preceding gates in a logic circuit. Finally, the logic-**1** output corresponds to the value $P^{\text{gate}}$ at time $t^{\text{gate}}$, which is set by the gate itself and therefore generally cannot be adjusted. Note that these definitions can only be made definitive in the framework of a particular application. In general studies of enzymatic gates it is sometimes convenient, for instance, to regard the slope of the time dependence of the signal, $dP(t)/dt$, which is nearly constant in the steady state regime typical for most enzymatic reactions, as the output.[16,32]

Ideally, our logic gate should only have chemical concentrations at logic-**0** or **1** values. However, due to noise in the system, concentrations not precisely corresponding to **0** or **1** are also possible. Let us define the (dimensionless) "logic" variables

$$x = I_1(0)/I_1^{\text{gate}}, \qquad (6)$$

$$y = I_2(0)/I_2^{\text{gate}}, \qquad (7)$$

$$z = P(t^{\text{gate}})/P^{\text{gate}}, \qquad (8)$$

in terms of which we can then consider the gate response function

$$z = F(x, y) = F(x, y; E(0), r, R, t^{\text{gate}}; ...). \qquad (9)$$

This function can then be studied for general $x, y, z$ ranging from **0** to **1** (and to values somewhat larger than **1**). The second expression in equation (9) emphasizes that the gate response function also depends on adjustable parameters (in this work, those displayed) that can in principle be varied to improve the gate performance, as well as on some parameters which are externally fixed (here this set could include $I_{1,2}^{\text{gate}}$, not displayed but indicated by "…").

Since the noise in general leads to some spread in the input variables around their logic-**0** and **1** values, the output logic variable $z$ also becomes not precisely determined. If we assume that the magnitudes of the deviations $\delta x$ and $\delta y$ are small and comparable to each other, then the resulting deviation, $\delta z$, for *smoothly varying* gate functions can be estimated as $\delta z \sim |\vec{\nabla} F| \delta x$, where the $|\vec{\nabla} F|$ is the magnitude of the gradient vector of the gate function at the appropriate logic point (**0,0**; **0,1**; **1,0**; **1,1**). This argument suggests that depending on the value of the largest of the four logic-input point gradients, the gate can amplify the noise level, $\delta z > \delta x, \delta y$, suppress it, $\delta z < \delta x, \delta y$, or keep it approximately constant, $\delta z \approx \delta x, \delta y$. The best-case scenario is, of course, the suppression of noise, but this is only possible when the response function has a "sigmoid" shape in both $x$ and $y$ variables, which has not been achieved for enzyme-based gates thus far. Since our enzyme (HRP) is expected not to possess such a "sigmoid" self-promoter characteristics for both the substrate and co-substrate, the best we could hope for in our biochemical system is signal propagation without noise amplification, i.e., $\delta z \approx \delta x, \delta y$, which, however, has been an elusive goal to achieve by direct optimization of the gate function dependence on adjustable parameters, as mentioned in the Introduction.

We point out that when the logic gate is part of an information-processing network, other considerations become important.[16] Specifically, "digital" noise which is a result of very-low-probability large fluctuations away from the logic values **0** and **1**, will become dominant for very large networks, and its buildup will have to be suppressed by network design and introduction of additional network elements, such as signal splitters and filters. These elements, especially filters, can also suppress the small-fluctuation "analog" noise discussed in the preceding paragraph. Thus, generally, control of noise buildup is possible by network, rather than only by individual gate optimization.[16]

Furthermore, even for individual gates the gradient-based consideration just advanced, is overly simplistic. Indeed, since the function $z = F(x, y)$ maps a two-variable input-noise distribution into a one-variable output-noise distribution, then for gate function shapes that are not very smooth, it is quite possible to have certain directions of large gradients but still obtain small $\delta z \approx \delta x, \delta y$. Our main finding in this work, detailed in Section 4, is that the regime with not very smooth response function can actually be advantageous for analog noise suppression. Thus, analysis in terms of the noise distribution is appropriate, as outlined in the rest of this section.

Following our recently developed approach,[15] to estimate noise amplification, we study the width of the output signal distribution, $\sigma_{\text{out}} = \sigma_z$, as a function of the width of the input noise distributions which will be assumed equal for simplicity and set to values $\sigma_{\text{in}} = \sigma_x = \sigma_y = 0.1$, which is a relatively large value based on the assumption that variations of the chemical concentrations in environments in which biochemical logic could be used in applications can be at least several percent. Furthermore, we will assume uncorrelated, Gaussian input noise distributions, with half-Gaussian for $x \geq 0$ at logic-**0**, and full Gaussian for $x$ at logic-**1**,

$$\text{logic-}\mathbf{0}: \quad G(x) = \frac{2e^{-x^2/\sigma_{\text{in}}^2}}{\sqrt{2\pi\sigma_{\text{in}}^2}}, \quad (10)$$

$$\text{logic-}\mathbf{1}: \quad G(x) = \frac{e^{-(x-1)^2/\sigma_{\text{in}}^2}}{\sqrt{2\pi\sigma_{\text{in}}^2}}, \quad (11)$$

and similarly for $y$. (Truncation to $x, y \geq 0$ for logic-**1** introduces negligible corrections and was neglected.) The output noise distribution width $\sigma_{\text{out}}$ is then estimated[15] by calculating

$$\text{logic-}\mathbf{0}: \quad \sigma_{\text{out}}^2 = \langle z^2 \rangle, \quad (12)$$

$$\text{logic-}\mathbf{1}: \quad \sigma_{\text{out}}^2 = \langle z^2 \rangle - \langle z \rangle^2, \quad (13)$$

where the moments such as $\langle z^2 \rangle$ of the gate response function $z = F(x, y)$ were computed[15] with respect to the input distribution $G(x)G(y)$.

This computation estimates the spread of the output signal near the respective logic value **0** or **1**, for the four logic input combinations **0**,**0**; **0**,**1**; **1**,**0** and **1**,**1**. In general one would want to have maximum of these spreads, $\sigma_{\text{out}}^{\max}$, to be as small as possible. In fact, for network scalability the actual value of the noise spread is not as important as the degree of noise *amplification* in each gate, measured by $\sigma_{\text{out}}^{\max}/\sigma_{\text{in}}$. As described earlier, the fact that the "gate machinery" enzyme, HRP, in our case is not expected to have self-promoter input(s) property ("sigmoid" gate-function shape), suggests that we can at best hope to have $\sigma_{\text{out}}^{\max}/\sigma_{\text{in}}$ values slightly over or equal to 1. Furthermore, earlier studies assuming "smooth" gate-function shapes indicate[15,16] that these can be optimized to at best achieve values slightly less than $\sigma_{\text{out}}^{\max}/\sigma_{\text{in}} = 1.2$. In the next section we demonstrate that $\sigma_{\text{out}}^{\max}/\sigma_{\text{in}} \approx 1$ is attainable in a new, not smooth, regime of gate-function shapes.

## 4. Results

### *4.1. Smooth gate-response shape with ABTS*

The biochemical logic gate mimicking Boolean **AND** was carried out in a buffered solution containing horseradish peroxidase (HRP) enzyme and other chemicals as described in Section 2.2. As illustrated in Scheme 1, biocatalytic oxidation of ABTS (or ferrocyanide, the latter reported in Subsection 4.3) resulted in the increased absorbance $\Delta A$, see Figure 1, of the solution which was measured at time $t^{\text{gate}}$ and converted to molar concentrations of oxidized ABTS (or ferricyanide). A change in the absorbance was not observed when only $H_2O_2$ was added to the solution (input signals **0**,**1**; note that the first digit corresponds to the logic input of the co-substrate: ABTS or ferrocyanide, while the second to hydrogen peroxide). Similarly, an absorbance change was not observed when only ABTS/ferrocyanide was added to the solution (inputs signals **1**,**0**). The presence of the both inputs in the solution (input signals **1**,**1**) resulted in the oxidation of two ABTS's (or ferrocyanides) for every $H_2O_2$ reduced.

The mapping out of the response function for the gate with ABTS is shown in Figure 2(A). In order to perform numerical optimization of the gate, we first fitted the experimental data by using the rate equations (3-5) at the reaction time 60 sec. This yields estimates of the rate constants $r$ and $R$. The resulting fitted response surface is presented in Figure 2(B). The fitted rate $r = 18\,\mu M^{-1} s^{-1}$ for the reaction step involving H$_2$O$_2$, is large, consistent with other published data.[27,30] The reaction step involving ABTS has a somewhat slower rate, $R = 5\,\mu M^{-1} s^{-1}$, consistent with literature estimates.[27] This rate is nevertheless also quite fast, comparable to the value of $r$.

Note that among the possible adjustable parameters available for gate-function optimization, $r$, $R$, $E(0)$, $t^{gate}$, the first two can be controlled by modifying the physical or chemical conditions in the system (as further commented on in Section 5) which is less straightforward that directly varying the last two parameters within reasonable ranges (to preserve the validity of the approximate rate equations used). Therefore, we first took the fitted reaction rates as fixed and computed the output noise level, as described in Section 3, for varying initial HRP concentration and reaction time. The results are shown in Figures 2(C,D), and we conclude that the "figure of merit" $\sigma_{out}^{max}/\sigma_{in}$ cannot be made smaller than $\approx 3$ for the scanned parameter ranges. This means that this ABTS-based **AND** gate significantly amplifies analog noise so that it is not good for utilization in even a very small information-processing network.

*4.2. Shape of the gate function and noise amplification*

The above conclusion is not surprising and was alluded to in the earlier discussion. Indeed, the shape of the response function for the **AND** gate with ABTS, Figure 2(A,B), is rather smooth and therefore one could work directly with gradients at the logic points rather than with the noise distributions. However, as found in earlier studies,[15,16] "balancing of gradients" at the logic points in order to make the largest of them as small as possible, is limited and can yield at best noise-amplification values slightly below 1.2. Even these values are not easy to achieve unless the enzyme concentration, for instance, is varied over a large range of a couple of orders of magnitude, which is not experimentally realistic.

However, HRP is known to take on a variety of co-substrates,[26] which offers an opportunity to have a large change in the rate *r*. Our numerical studies of the desired rate ranges, based on the specific rate equations (3-5), yielded as interesting conclusion (which is generally applicable and not limited to this particular kinetics). Specifically, we found that in terms of the noise distributions, one can achieve values of $\sigma_{out}^{max}/\sigma_{in}$ rather close to 1 in the regime of large imbalance of the rates, here $r \gg R$. Of course, one still has to find the proper values of the other system parameters, but these turn out experimentally reasonable and are actually realized in the next subsection.

This result is at first surprising, because in such regimes the gate response surface is not "smooth" but has a (rounded) "ridge" and directions of fast variation at the logic points, suggesting large gradients, as illustrated in the next subsection; see specifically Figure 3. However, due to the fact that we have a map of two-variable distributions into one-variable distributions, having large slopes but along limited angular ranges of directions away from the logic points, mathematically does not lead to large spread in the signal distribution, allowing for gate operation without significant analog noise amplification. Essentially, large slopes have "small weight" in this case and therefore do not result in broadening of the output distribution.

*4.3. Low-noise gate shape with ferrocyanide*

The experimental and fitted response functions for the **AND** gate with ferrocyanide as an input, are shown in Figure 3(A-B), where one can see that these surfaces can be roughly represented by two intersecting planes. By comparing the fitted surfaces, Figures 2(B) and 3(B), with the experimental ones for ABTS and ferrocyanide substrates, Figures 2(A) and 3(A), respectively, we concluded that our rate equations (3-5) in fact describe the system with ferrocyanide much more accurately than for the case of ABTS. Note that we actually also performed simulations (not reported here) using a more complex system of chemical reactions for ABTS oxidation[27] with three adjustable rates and found that the resulting fit is not surprisingly (for more parameters) in a better agreement with the experimental data. However, the rate constants extracted from the two models are still of the same order of magnitude. Therefore, we prefer to use the same rate equations to describe the kinetic properties of both systems.

The reaction rates obtained from the fitting of the experimental data are $r = 17\,\mu M^{-1} s^{-1}$ and $R = 32 \times 10^{-3}\,\mu M^{-1} s^{-1}$, consistent with previous works.[26] Note that the value of $R$ in the case of ferrocyanide is about 170 times smaller than for ABTS. From Figure 3(C,D) one can see that the width of the output noise distribution shown achieves a marked minimum at which $\sigma_{out}^{max}/\sigma_{in} \approx 1$ for properly selected values of the two parameters $E(0)$, $t^{gate} = E^{min}, t^{min}$, and that $E^{min}(t^{min})$ is a monotonically decreasing function. Figure 3(D) also suggests that our experimentally convenient for otherwise randomly selected values of $E(0) = 0.5\,\mu M$ and $t^{gate} = 60$ sec, correspond to $\sigma_{out}^{max}/\sigma_{in} \approx 2$. The latter value is already better than that for ABTS, but it should be further optimized if this gate is to be used as part of a network.

The important feature in the new identified regime of the non-smooth gate shapes, is that the adjustment of the concentrations required to reach the optimal functioning is quite reasonable and does not involve changes by orders of magnitude. As pointed out in Section 3, some parameters might be fixed by the gate's surroundings (network). Let us consider, for instance, $I_2^{gate}$ — the concentration of the input $H_2O_2$. If variation of this parameter is possible in a particular application, then its adjustment required to achieve optimal gate functioning (with all the other parameters fixed) is also quite reasonable. As shown in Figure 4, reduction of the concentration from our original value of 250 μM to 150 μM yields $\sigma_{out}^{max}/\sigma_{in}$ very close to 1.

## 5. Conclusion

We comment that in all the $\sigma_{out}^{max}/\sigma_{in} \approx 1$ situations, the "ridge" (cf. Figure 3) in the response function extends directly from the logic-**0,0** point to logic-**1,1**, resulting in the smallest possible noise amplification of about 5% at all logic points (i.e., $\sigma_{out}^{max}$ of order $1.05\sigma_{in}$), see Figure 4. In fact, in our experiments essentially the same value of $\sigma_{\text{out}}^{\max}/\sigma_{\text{in}} \approx 1$ could be obtained in several ways.

Our "optimization" actually started by selecting the pH, as described here. For the logic gate with ferrocyanide as the co-substrate, the rate $r$ strongly depends on pH.[26] Increase of pH leads to the shift of the "ridge" towards the **1,1** logic point, thereby expanding the range of possible $I_2^{\text{gate}}$ values. This is the reason why our final experiments with ferrocyanide (Figures 3 and 4) were carried out with pH = 6. Our preliminary experiments for ferrocyanide (not detailed here) were with pH = 5, the same as for ABTS. This, however, resulted in too steep a slope at the **1,0** logic point, cf. Figure 3(A,B). The larger value of pH = 6 was thus selected, somewhat by trial and error, to resolve this difficulty. This approach generally illustrates the idea of controlling the reaction rates by changing chemical (or physical) system parameters, and the fact that such control is difficult to predict systematically.

However, to obtain $\sigma_{\text{out}}^{\max}/\sigma_{\text{in}} \approx 1$ we have to vary other parameters as well. One possibility is to adjust the value of $I_2^{\text{gate}}$ while keeping the rest of the system parameters intact, as discussed above in connection with Figure 4. We also carried out experiments (not detailed here) that involved an adjustment of the "gate machinery" by taking the reaction time to a value close to the pair of values $(E^{\min}, t^{\min})$, without changing the initial HRP concentration: $E^{\min} = E(0)$, see Figure 3. Specifically, we checked that data taken for $E(0) = 0.05\,\mu\text{M}$, $t^{\text{gate}} = 120\,\text{sec}$ yield a "planar" response surface, with the "ridge" extending from logic-**0,0** towards logic-**1,1**, as predicted theoretically.

To put this observation, and our results, in a broader context, we offer the following summary. We note that typical (bio)chemical systems with two inputs and one output, mimicking the **AND** logic, correspond to the gate-function surface of shape schematically drawn in Figure 5(A), in which, unlike Figures 3 and 4, we use the "logic" variables scaled to the interval form 0 to 1. Indeed, in Figure 5(A) the output is linear in each of the chemicals when their small concentration is the limiting factor, but reaches saturation for larger input values. Earlier studies[15,16] found that modification of the system parameters which does not dramatically change this shape, can yield at best values not much below 1.2 for the ratio $\sigma_{\text{out}}^{\max} / \sigma_{\text{in}}$.

Our findings in this work suggest that this value can be made smaller, close to 1, for the shape schematically shown in Figure 5(B). This surface has some slopes larger than 1, but these have small "weights," as explained earlier. In addition to this new gate-function shape, there is also an approach, Figure 5(C), to have an input with self-promoter properties at small concentrations. While there is a possibility to identify allosteric enzymes with such behavior, this option, which will allow getting the ratio $\sigma_{\text{out}}^{\max} / \sigma_{\text{in}}$ down to 1 for proper parameter values, has thus far not been experimentally realized in the context of biocomputing gates.

Finally, Figure 5(D) presents the "ideal" case of self-promoter behavior in both inputs, which truly generalizes the single-variable "sigmoid" response, and which can yield actual analog noise suppression ($\sigma_{\text{out}}^{\max} / \sigma_{\text{in}} < 1$). While such mechanisms are encountered[29] in complicated natural processes, they are unlikely to be realized for a single gate with one or few enzymes as the "machinery," though they might be possible if additional processes which are not part of the enzymatic gate-function are added. All these options are presently theoretical but will be subjects of future work.

**Acknowledgements**

We gratefully acknowledge support of our research programs by the National Science Foundation (grants CCF-0726698 and DMR-0706209), by the Office of Naval Research (grant N00014-08-1-1202), and by the Semiconductor Research Corporation (award 2008-RJ-1839G). GS acknowledges Wallace H. Coulter Scholarship at Clarkson University.

# Scheme and Figures

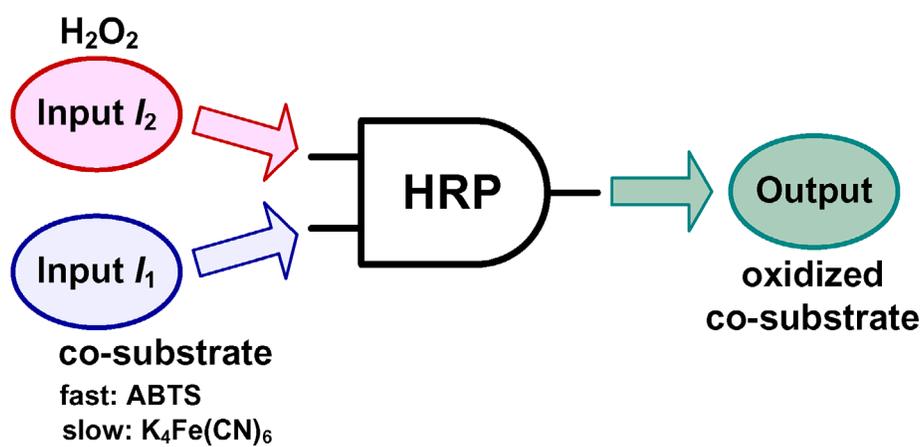

**Scheme 1.** Single-enzyme **AND** logic gate.

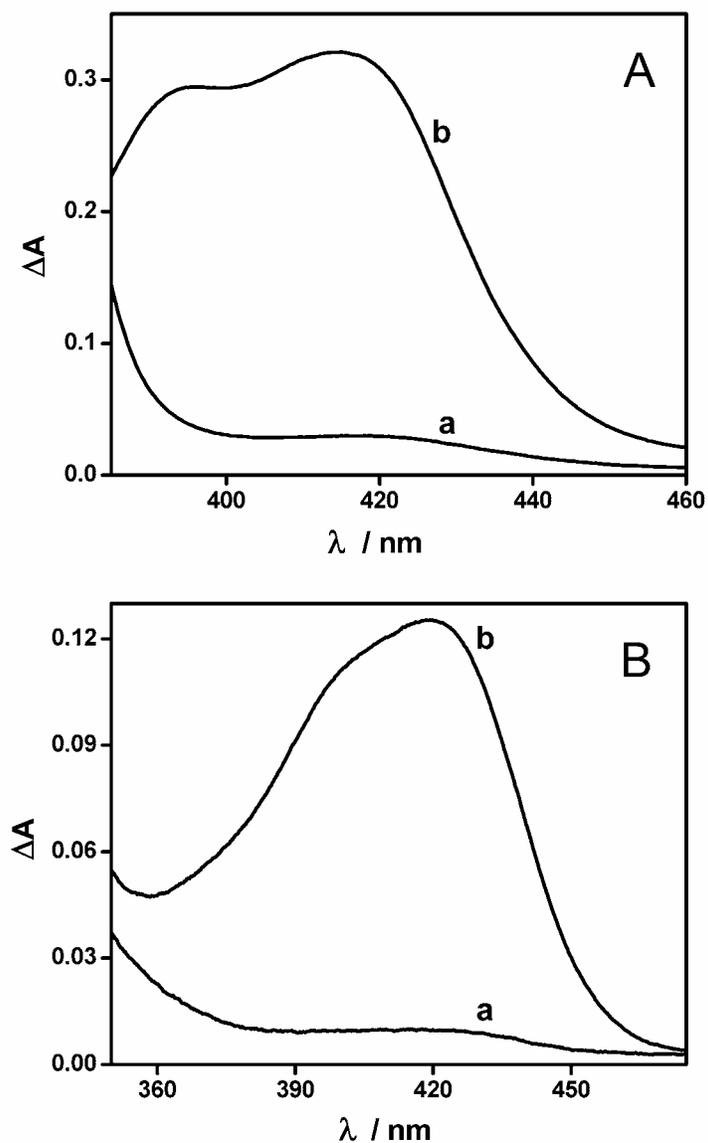

**Figure 1.** (A) Typical spectral features of the **AND** gate after 60 sec reaction with HRP, 0.5 nM, and co-substrate ABTS, 0.5 mM, in solution: (a) without the addition of $H_2O_2$ (**1,0**); (b) after the addition of $H_2O_2$, 50 μM (**1,1**). (B) Spectral features of the **AND** gate after 60 sec reaction with HRP, 0.05 μM, and with ferrocyanide, 1.5 mM, as a co-substrate in solution (a) without the addition of $H_2O_2$ (**1,0**); (b) after the addition of $H_2O_2$, 0.25 mM, (**1,1**).

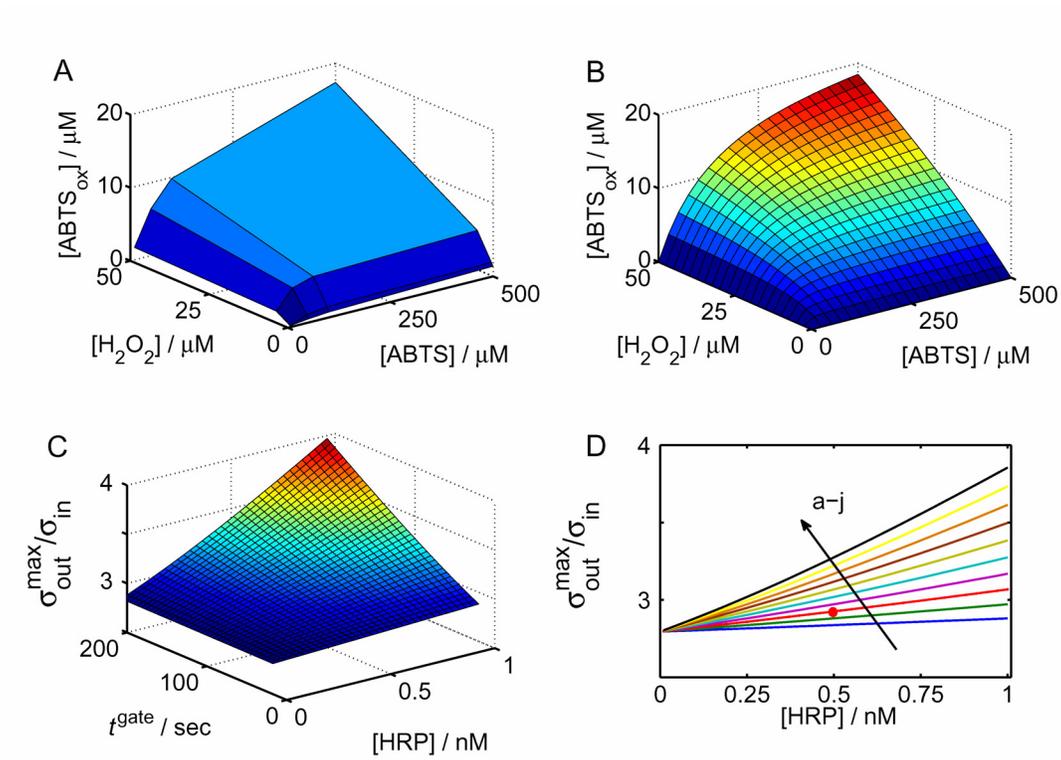

**Figure 2:** (A) Measured and (B) numerically fitted response surface for the enzymatic logic gate with ABTS as one of the inputs. (C) Surface plot of the gate function quality measure, $\sigma_{out}^{max}/\sigma_{in}$, as a function of the enzyme concentration and reaction time. (D) Dependence of $\sigma_{out}^{max}/\sigma_{in}$ on HRP concentration for different reaction times. Curves labeled (a–j) in the order indicted by the arrow, correspond to $t^{gate} = 20, 40, 60, \ldots, 200\,\text{sec}$. The red dot marks our experimental conditions: $[\text{HRP}](t=0) = 0.5\,\text{nM}$, $t^{gate} = 60\,\text{sec}$.

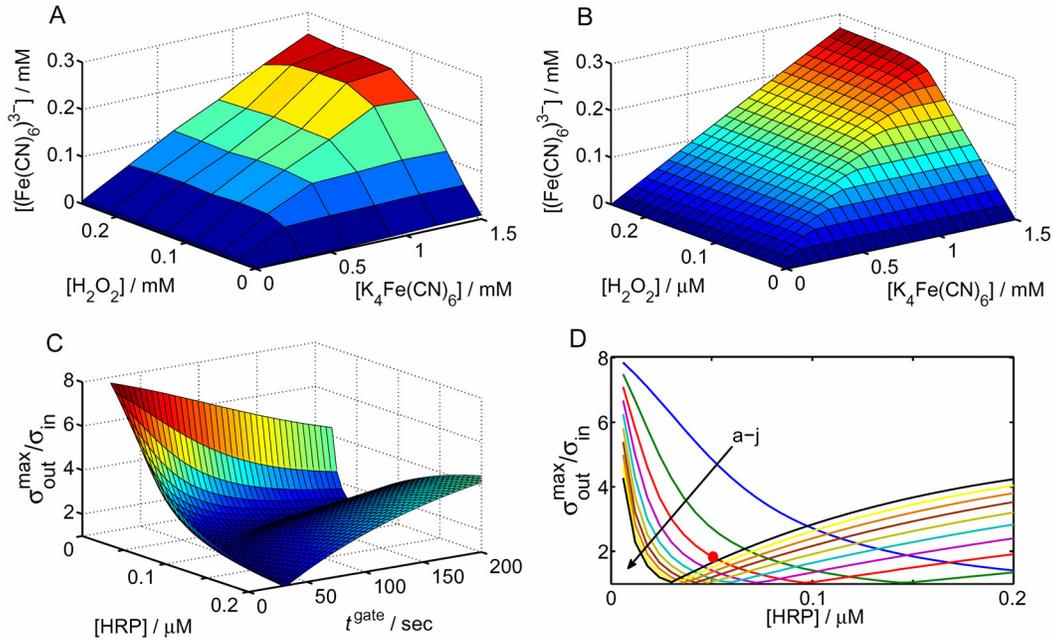

**Figure 3:** (A) Measured and (B) numerically fitted response surface for the enzymatic logic gate with ferrocyanide as one of the inputs. (C) Surface plot of the gate function quality measure, $\sigma_{out}^{max}/\sigma_{in}$, as a function of the enzyme concentration and reaction time. (D) Dependence of $\sigma_{out}^{max}/\sigma_{in}$ on HRP concentration for different reaction times. Curves labeled (a–j) in the order indicted by the arrow, correspond to $t^{gate} = 20, 40, 60, \ldots, 200\,\text{sec}$. The red dot marks our experimental conditions: $[\text{HRP}](t=0) = 0.05\,\mu\text{M}$, $t^{gate} = 60\,\text{sec}$.

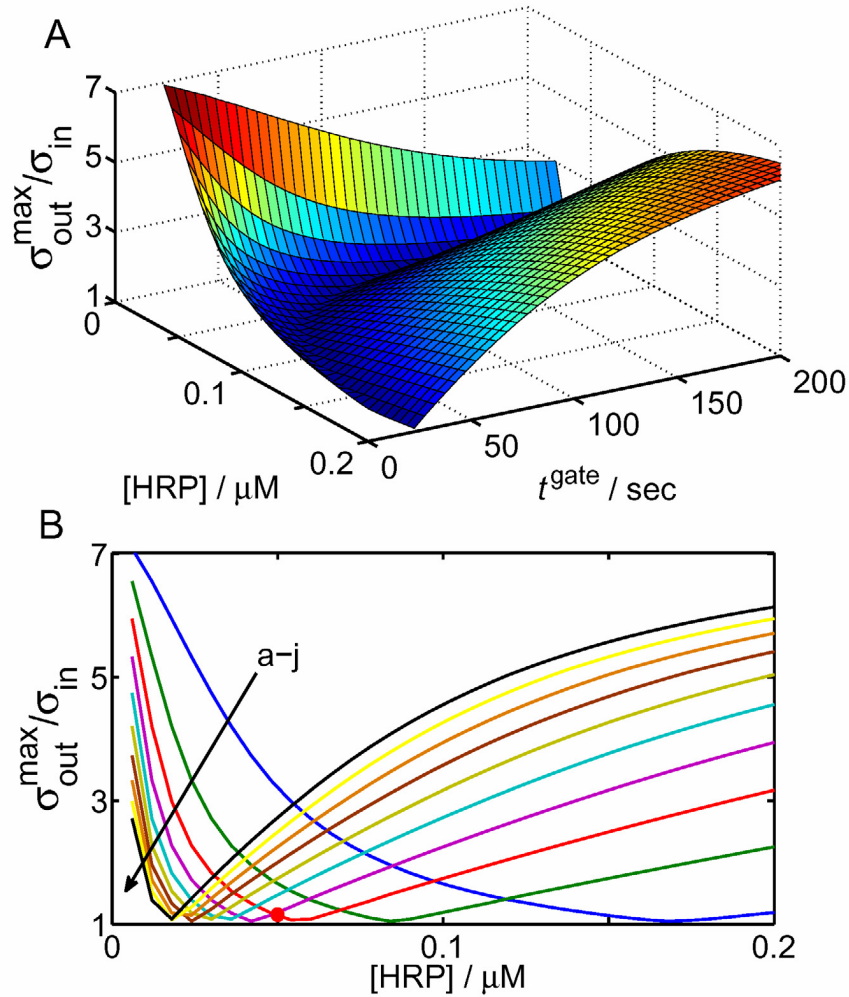

**Figure 4:** (A,B) — the same as in Figure 3(C,D), but for initial (input) H$_2$O$_2$ concentration $I_2^{\text{gate}} = 150$ μM: (A) Surface plot of the gate function quality measure, $\sigma_{\text{out}}^{\max}/\sigma_{\text{in}}$, as a function of the enzyme concentration and reaction time. (D) Dependence of $\sigma_{\text{out}}^{\max}/\sigma_{\text{in}}$ on HRP concentration for different reaction times. Curves labeled (a–j) in the order indicted by the arrow, correspond to $t^{\text{gate}} = 20, 40, 60, \ldots, 200\,\text{sec}$. The red dot marks our experimental conditions: [HRP]$(t=0) = 0.05\,\mu\text{M}$, $t^{\text{gate}} = 60\,\text{sec}$.

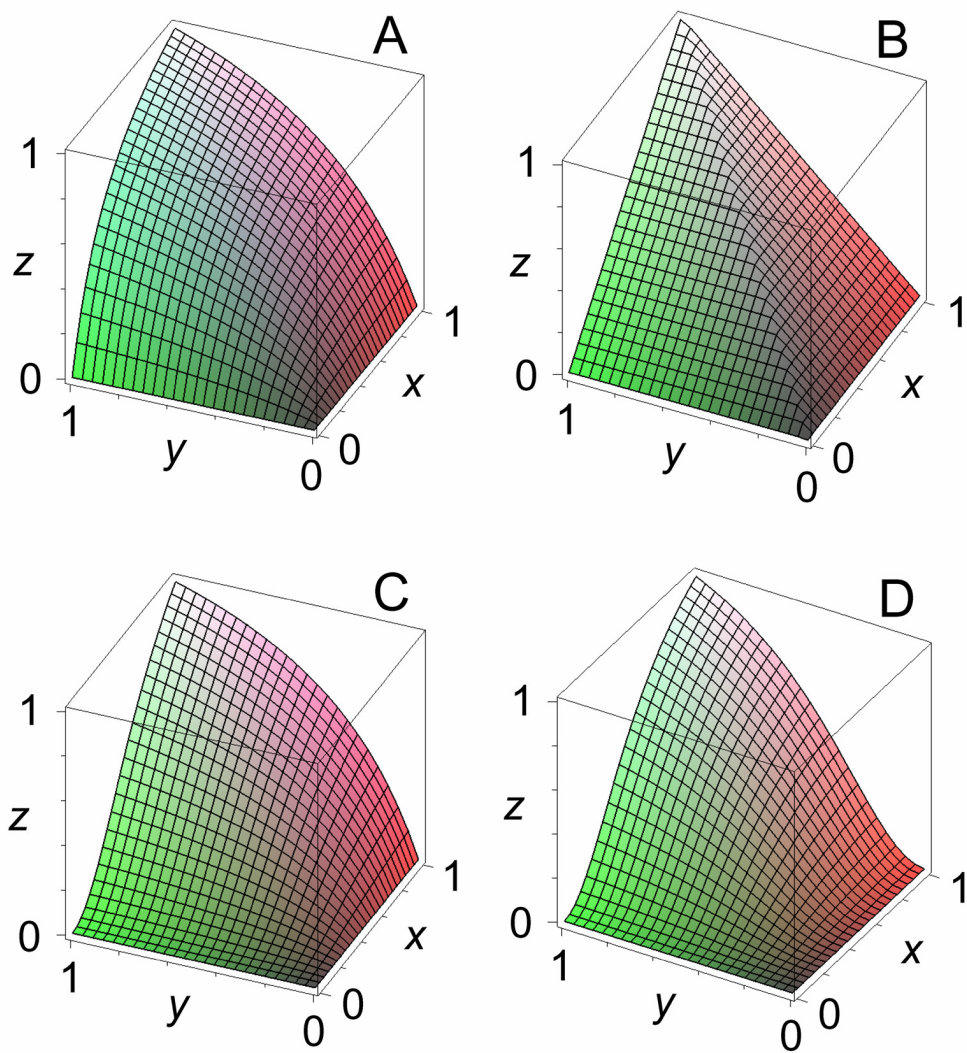

**Figure 5:** Schematic of the gate-response functions encountered in this work. (A) The "smooth" response obtained with ABTS as one of the inputs. This case is common but difficult to optimize at the level of a single gate function. (B) The "ridged" shape (here shown very close to the optimized, symmetric ridge position) obtained with ferrocyanide as an input. Our main finding in this work is that such response is relatively easy to optimize. (C) Response for the case of one of the inputs having the "self-promoter" property. While not experimentally studied, such systems, when properly optimized, can lead to logic gates without noise amplification. (D) The ideal case of "self-promoter" response in both inputs, which, while not common for simple enzymatic reactions, is encountered in natural systems and with proper parameter choices (which is not the case shown here), can yield actual noise suppression.